\documentclass{article}

\begin{document}

\title{Report on A5. Computer Methods}

\author{Beverly K. Berger \\
Physics Division, National Science Foundation,
Arlington, VA  22230 USA\\ 
E-mail: bberger@nsf.gov}


\maketitle

\begin{abstract}
Session A5 on numerical methods contained talks on colliding black holes,
critical phenomena, investigation of singularities, and computer algebra.
\end{abstract}

\section{Introduction}
In this Chapter,
I shall describe the current status of computer simulation and algebraic
computing for general relativity and gravitation.
Most of the effort, both as represented in this Workshop and world-wide,
is focused on ``numerical relativity,'' the name usually given  to
large scale simulations of isolated gravitational
systems. Elsewhere in these proceedings is the contribution by Lehner
based on his plenary talk on the status of the colliding black hole (CBH)
problem. This problem is extremely difficult but urgently
needs to be solved. The urgency arises because one desired output of CBH simulations
is a set of gravitational wave forms that can be used by the ground based
gravitational wave detectors (LIGO, VIRGO, GEO600, and TAMA) now coming
into operation. The difficulties arise on many fronts. These include (1)
the computational spacetime grid is not the physical spacetime grid; (2)
CBH spacetimes are singular within the horizons of the black holes. Note
that simulations cannot be followed (much) past the time at which NaN
(not a number) or INF (infinity) is recorded as a value anywhere in the
computational grid; (3) Setting up initial data is highly nontrivial. In
the process of solving the constraints, the connection between the actual
initial system and the desired physical initial state is lost. For
example, the ``no incoming radiation'' boundary condition is problematic;
(4) Existing 3D CBH codes which have made considerable progress in dealing
with (1) and (2) are plagued by instabilities which cause them to crash
long before any physically relevant results are obtained. Note that a code
that runs long enough can overcome the problem of radiation in the
initial data by evolving until after it dissipates; (5) Identification of
outgoing gravitational radiation is also nontrivial because radiation is
a small effect thus requiring accurate simulations for a credible
extraction and because many computational grids must terminate before the
radiation zone is reached. All of these issues are under intensive
investigation and were discussed in this Workshop. Similar techniques
have been developed to study the coalescence of neutron star binaries.
Such simulations do not have to face the issues of singularity and
horizon but do have the complication of matter.

The travails of the CBH simulations should not be taken to mean that
all computer simulations in general relativity and gravitation have not yet
been completely successful. In fact, rather the opposite is true. Most spectacular,
of course, was the discovery by Choptuik of critical phenomena in the
collapse of a spherically symmetric self-gravitating scalar field.
Careful numerical techniques (including adaptive mesh refinement)
allowed study of the critical system dividing scalar field configurations
which eventually disperse from those which eventually form BHs. Scaling
typical of that found in other phase transitions was observed. The
critical solution itself (a zero mass naked singularity) displays
periodic self-similarity --- a phenomenon unknown outside gravitating
systems. Choptuik's original discovery sparked a large number of further
and ongoing numerical and analytic studies of critical phenomena in
gravitational collapse. Some of these were discussed in this Workshop.

Another area where there has been some success has been the numerical
investigation of cosmological singularities. While an actual singularity cannot be
handled on the computer, the regular behavior in the approach to it may be
studied. If, asymptotically as the singularity is approached, some types
of terms in Einstein's equations dominate over others, it might be
possible to characterize the approach to the singularity in a simple way.
This was argued heuristically long ago by Belinskii, Lifshitz, and
Khalatnikov (BKL) with their predictions of local Mixmaster dynamics
(LMD) recently supported in the cosmological context with numerical
experiment and, in some cases, mathematical proof. In fact, the synergy
between the mathematical and numerical approaches in this area has been
useful in understanding the relevant phenomenology. 

Originally, this Workshop was designed to feature algebraic computing as
well as computer simulation. However, the co-organizer, Jim Skea, was
unable to attend. In the end, there was only a single talk on this
subject where considerable progress was described toward the goal of
automated classification of exact or perturbative solutions of Einstein's
equations.

In the rest of this Chapter, the highlights of the talks in each area
will be given. For more details please see the indicated references.

\section{Colliding Black Holes and Neutron Stars}
Lousto presented an overview of the Lazarus Project to
combine 3D simulations and perturbation approaches in an ambitious attempt to follow a
BH collision and extract waveforms.\cite{baker01a} This required the development
of techniques to convert the coordinates suitable to the 3D simulation to those
appropriate for the single black hole close-limit approximation. 
These simulations were checked by
interchanging the use of 3D simulation and perturbation 
where both were valid. In addition, the method was applied to a single Kerr BH
where computed spurious radiation indicates the degree of error. This approach has
made it possible to compute the first complete waveforms covering the post-orbital
dynamics of a binary black hole system.

A number of talks were given on various techniques needed for standard (i.e. $3 +
1$ formulation) 3D codes. It is currently believed that an important source of
instability in 3D codes is constraint violating modes. Einstein's equations may be
written in an infinite number of ways by adding arbitrary multiples of the constraints
to the dynamical equations of motion. The goal is to identify a formulation which will
drive the system to the constraint hypersurface and thus eliminate these
instabilities. One such approach was discussed by Shinkai.\cite{shinkai01}

Pollney discussed work (with AEI collaborators) on BH excision and gauge
techniques.\cite{alcubierre01} They describe recent progress in an implementation
of excision for 3D black holes along with a set of gauge conditions which respond
naturally to spacetime dynamics. Through the combined use of these
techniques, they are able to produce accurate and long-lived evolutions
of highly distorted, rotating black hole spacetimes. 
Accurate waveforms can be extracted.

Diener presented the results of the head-on CBH code test for 3D simulations. Koppitz
reported on preliminary work in the construction of initial data sets. The objective
is to consider thin sandwich, Kerr-Schild, and post-Newtonian approximation initial
data using various implementations. 

Bondarescu described how one might visualize BH horizons as they evolve in
CBH simulations.\cite{bondarescu01}  The 3D codes 
allow one to locate and track the evolution of apparent and event
horizons in the coordinates of the simulation.  One typically has little information
about their real geometry. Previous studies in axisymmetric spacetimes have visualized
horizons via an embedding in flat space. A new method which goes beyond
axisymmetry was discussed. The method correctly reproduces
known results and, for the first time, has allowed construction of embeddings of
non-axisymmetric, distorted black holes.

Ashtekar described a method to extract physics from strong field regimes in numerical
simulations using the isolated horizon (IH) framework.\cite{ashtekar00b} 
If the IH (but not the spacetime) is axisymmetric, e.g., then, knowing just the
intrinsic metric and one Newman-Penrose spin coefficient, the IH 
framework enables one to calculate the angular momentum and
mass of the final black hole. These quantities refer only to the black hole in
equilibrium and do not include the angular momentum and mass in
matter fields or radiation outside the IH and thus could be found numerically. 

Grandcl\'ement described a new approach to binary black hole evolution prior to the
plunge phase.\cite{GrandGB01}  One important feature of this phase of the
evolution is the location of the innermost stable circular orbit (ISCO). This has
been calculated both using post-Newtonian approximations and numerically from
quasi-static equilibria with different results. If two orbiting 
bodies are far apart, the time-scale associated with the gravitational radiation is
much longer than the orbital period allowing one to assume that the two black holes
are on exact circular orbits. 
Using multi-domain spectral methods, a sequence of two identical 
corotating black holes is computed.
The ISCO appears as a turning point in the total energy and angular momentum
curves. Its position agrees
well (for the first time) with the post-Newtonian values.

Several talks were given on characteristic codes. For a review of this approach,
see [\cite{winicour00}]. Bishop described the initial value problem for a 3D
characteristic code. The code is robust and stable and has been
extended to include matter,\cite{mat} although in a simple way that would not
handle shocks. The code cannot evolve a whole spacetime, but requires data on an
inner world tube. For this reason, it is important to identify initial data of the
required type to represent a physical situation with strong gravitational fields.

D'Inverno and Pollney discussed lower
dimensional code tests for a Cauchy characteristic matching (CCM) code. See (e.g.)
[\cite{dinverno00}]. The advantages of such a code are that it dispenses with an
outer boundary condition and (since it uses a compactified coordinate) can yield
global solutions in which gravitational waves
can be identified unambiguously at future null infinity. Recently, a master vacuum
axisymmetric CCM code has been completed. The main motivation for this work is to
construct a three dimensional code possessing the characteristic,
injection and extraction modules (now present in the 2D code) which can be attached to
existing interior codes based on a finite grid.

A completely different approach involves a 3D code based on the conformal field
equations developed by Friedrich. This work was presented by
Husa. He discussed the evolution of linear and (mildly) nonlinear gravitational waves.
This approach aims at the numerical computation of
the global structure of generic isolated systems in GR using
well defined and unambiguous algorithms even out to and
beyond future null infinity.
The basic ideas and algorithms were developed by 
H\"ubner and Weaver in recent years and implemented
by  H\"ubner.
H\"ubner\cite{PeterIV} obtained
the complete future of (the physical part of) the initial slice, illustrating
a theorem by Friedrich, which states that
for sufficiently weak initial data a regular point $i^+$ exists.
However Husa found that for higher amplitudes the gauge chosen in by H\"ubner
results in code crashes which can be cured
by some ad-hoc modification of the lapse. For still stronger (but still only
mildly nonlinear) data, evolution up to $i^+$ would require the tuning of
too many parameters to compensate for the lack of symmetry. 
Evolutions of axially symmetric initial data 
modeled in the spirit of ``Brill waves'', and
evolutions of Minkowski space in a static gauge were discussed.
While the former evolution is numerically stable and quite robust, the latter
exhibit instabilities which seem to be rooted in exponentially growing constraint
violating modes inherent in the analytical formulation.

Siebel presented the results of a study using the characteristic approach to
identify gravitational radiation. The implementation follows the characteristic
initial value formulation of general relativity based on Bondi's radiative
metric.
The study focuses on the evolution of 
neutron stars modeled by a $N=1$ polytrope with fully relativistic
hydrodynamics.\cite{SFP01} 
The code was used
to study spherical relativistic stars with a 
scalar field to model the gravitational degrees of freedom.
Depending on the specific neutron star model, the scalar field 
either induces oscillations of the star or makes 
it collapse to a black hole. The extracted frequencies of the oscillations agree very
well with linear studies, except, as expected, at the threshold of black hole
formation.   The code was also applied to evolutions 
of neutron stars in axisymmetry and to their interaction with gravitational 
waves.

Stark (with Lun) considered an alternative $3 + 1$ scheme as applied to
spherical perfect fluid collapse. The scheme utilizes the constraint equations
contained in the Einstein field equations with those in the Bianchi identities to
determine the 3-curvature tensor and the 3-covariant derivatives of the
extrinsic curvature of the hypersurface.  This yields a
hyperbolic system of equations for the evolution of the gravitational
potentials and hydrodynamical quantities.
A numerical implementation of this system was used to model the Oppenheimer-Volkoff
solution and the  gravitational collapse of an initially homogeneous dust sphere to a
BH.  

Moreschi discussed modeling the collision of black holes with angular momentum with
Robinson-Trautman (RT) spacetimes.\cite{moreschi01} These spacetimes contain
purely outgoing gravitational radiation and decay to leave a Schwarzschild-like
horizon. These RT spacetimes are natural candidates to 
study systems settling to a single black hole and may be used as background
spacetimes for a perturbation treatment.
Previous studies of this type
give excellent agreement with numerical relativity results for the head-on
collision, zero spin case. They are then extended to include angular momentum. 

Reula presented results using the collapse of spherically
symmetric self-gravitating scalar fields as a model problem to study numerical
instabilities.\cite{iriondo01}
There are instabilities  present
in this simple problem, and furthermore there are strong indications
that, more generally, the instabilities are  predominantly
due to a longitudinal or Newtonian mode, namely the only one present in the
spherically symmetric case.
Free and constrained evolutions are compared. 
In  [\cite{Sch-Baum_1}] a symmetric
hyperbolic formulation is used, and certain freedom, still available in that
setting, is used to suppress the main instability found there,  
to allow a stable propagation. 
However, their discretization scheme has first order numerical dissipation 
and the boundary-value problem for the system is not known to be well posed. 
Reula discussed an improved system that overcomes these problems,
preserves the constraints, and reproduces several known results.

Gentle applied Regge calculus methods to the evolution of Brill waves.  
This is the first significant test of Regge calculus in a highly
dynamic setting.  An axisymmetric lattice is obtain by
triangulating and collapsing a hypercubic grid which is aligned with a
cylindrical polar coordinate system (see [\cite{gentle99}]
and references therein).

Neutron stars might act as a source of gravitational waves. Sperhake presented a
new numerical approach to nonlinear oscillations in neutron
stars.\cite{sperhake01} This approach rewrites the equations to cancel out the
background terms in the numerical evolution of the perturbations.  A comparison of
the resulting perturbative code with a ``standard'' non-perturbative method was
made.  The perturbative scheme reproduces the expected harmonic oscillations with
high accuracy, while the non-perturbative scheme produced spurious behavior. The
new method is then used to accurately evolve initial data of arbitrary amplitude to
investigate non-linear coupling of eigenmodes for a simplified neutron star model.

Hawley (with Choptuik) explore the genericity of initial data for
multi-scalar stars. 
They consider a class of general relativistic soliton-like solutions
composed of multiple minimally coupled, massive, real scalar fields
without self-interaction.

\section{Numerical Investigation of Singularities}
For reviews of critical phenomena see [\cite{gundlach99a,berger98}]. 
Gundlach described
what happens to the critical behavior of the gravitational collapse of 
a perfect fluid
when there are departures from spherical symmetry. 
Does the type II critical behavior in the gravitational collapse of a $p = k \rho$
perfect fluid persist beyond the restriction to
spherical symmetry? It was found that for $1/9<k<0.49$, all non-spherical
perturbations decay. For $0.49<k<1$, there are one or several
growing modes in the polar perturbations. For $ 0<k<1/9$, there is exactly one
non-spherical growing mode. It is an $l=1$ axial mode, linked to
infinitesimal rotation.
BH mass and BH angular momentum become a function of
2-parameter families of initial data, chosen so that one of the
parameters is linked to rotation (and thus can be tuned to zero by axisymmetry).
At the BH threshold they depend on the initial data
through one universal function of one variable,a combination of these two
parameters. The formula obtained
is completely analogous to the dependence of the magnetization of a ferromagnet as
a function of temperature and external magnetic field near its critical point.

Aichelburg presented results (including those of Lechner) on a new transition between
discrete and continuous self-similarity in collapsing $SU(2)$ $\sigma$-models
parametrized by a dimensionless coupling constant.\cite{thornburg01}    
In the intermediate range,
 a competition between CSS and DSS solutions gives rise to new phenomena
in the dynamical evolution of a critical search: the appearance of
episodes of approximate CSS, where repeatedly the evolution approaches
and departs from CSS before leading to black hole formation or dispersion.
This picture is supported by the explicit numerical construction of the 
CSS and DSS solutions and its comparison with critical time evolutions.

Szpak studied critical behavior in a model nonlinear wave problem.\cite{Mth} The
motivation was to study the existence of an  universal intermediate attractor in
the dynamics at the threshold of forming  singularity. 

Hobill (with Webster) presented studies of trapped surface formation in Brill wave
evolution to see if a naked singularity could form. The oblateness or prolateness
of the system is determined by measuring the ratio of polar circumference
to equatorial circumference for a surface with constant logarithmic
radial coordinate.  It is found that for initial data with positive
amplitude the systems are prolate and for negative amplitude the
systems are oblate. It is seen that,
as the Brill wave amplitude is increased (thereby decreasing the
radius of the outermost minimal surface), the location of the outermost trapped
surface eventually coincides with the minimal surface thereby preventing the
formation of a naked singularity. This behavior is in agreement with recent
results of Garfinkle and Duncan.\cite{GarfinkleDuncan} 

I presented a summary of our (with D. Garfinkle, J. Isenberg, V. Moncrief, M.
Weaver) program of investigation of collapsing spatially inhomogeneous
cosmologies.\cite{berger98} Significant recent progress has been made in rigorous
studies of generic collapse in cases where one expects the approach to the
singularity to be asymptotically velocity term dominated.\cite{andersson} However,
one expects that an even larger class of spacetimes exhibits local
Mixmaster dynamics (LMD) in the vicinity of the singularity. While numerical
evidence for LMD in spatially inhomogeneous spacetimes with
$T^2$ and $U(1)$ symmetries is compelling, rigorous results for
spatially inhomogeneous cosmological spacetimes with LMD do not exist.
New mathematical methods to handle LMD are needed.

\section{Algebraic Computing: Invariant Classification in Maple}
D'Inverno provided a summary of the program to classify solutions to Einstein's
equations automatically.\cite{pollney00a} Recent
developments are in collaboration with Pollney, Skea, Ara\'ujo, de Albuquerque,
and Roveda. A short review of the equivalence problem in general relativity was
presented which included a description of the computer database of exact solutions
whose central site is located at  http://www.jim.dft.uerj.br. The database
contains over 200  solutions of Einstein's equations which have been classified
with CLASSI, an extension of the general relativity system called  SHEEP. In the
next stage of this collaborative  project attention has been focused on setting up
a database which can easily be accessed and  updated by the user community. The
choice of platform is the general relativity package
GRTensor which is an applications package of the computer algebra system
Maple. The underlying algorithm is based on the Cartan-Karlhede invariant
classification of geometries. In particular, new algorithms have been obtained for 
putting the Weyl spinor, Ricci spinor and general spinors into standard 
form and the derivative operators needed in the classification algorithm 
and the behavior of symmetric spinors under frame rotations have been
investigated. 
More recent 
work was reported on a refinement of the JMS (Joly-MacCallum-Seixas)
algorithm for completely determining the Segre type of a Ricci spinor.
To date 30 of the 45 possible sub-cases have been completed and the
algorithms have been implemented in GRTensor. In addition, the status of
the work on obtaining the isometry group of a space-time (rather than
simply its dimension as in the current database) was reported. It is
currently possible to obtain the isometry group for spacetimes admitting a
G2, G3 and some G4 symmetries. This may require the user to input limits
on parameters and coordinates appearing in the line element, to overcome
an ambiguity in determining the signature of a matrix occurring in the
Bianchi classification. The goal is to complete the G4 case and extend the
work to the G5, G6 and G7 cases.
Work was also reported on the Maple package SPIDOR, which has been
developed for obtaining spinor contractions. An example of its application
is its ability to automatically generate the spinor curvature invariants
built out of the Weyl spinor, its complex conjugate and the Ricci spinor
up to the 20th order on a laptop. This package may be useful in relation 
to formalisms which have been developed to address particular issues in 
exact solutions.

\end{document}